\documentclass[12pt]{iopart}
\usepackage{iopams}
\usepackage{tikz}
\usetikzlibrary{calc}
\usepackage{ytableau}
\usepackage{tikz}
\usepackage[latin1]{inputenc}
\usetikzlibrary{calc,arrows,positioning}
\usetikzlibrary{trees}
\usetikzlibrary{decorations.pathmorphing}
\usetikzlibrary{decorations.markings}
\newcommand{\ben}{\begin{equation}}
\newcommand{\bal}{\begin{align}}
\newcommand{\een}{\end{equation}}
\newcommand{\eal}{\end{align}}
\newcommand{\bea}{\begin{eqnarray}}
\newcommand{\eea}{\end{eqnarray}}
\newcommand{\nn}{\nonumber\\ }

\newcommand{\vp}{{\varphi}}
\newcommand{\vpd}{\varphi^\dagger}

\newcommand{\qq}{\qquad\qquad}
\newcommand{\QQ}{\qquad\qquad\qquad\qquad}

\newcommand{\cA}{{\cal A}}

\newcommand{\hfb}{{\hfill\break}}
\newcommand{\sub}[1]{_{\stackrel{}{#1}}}

\input amssym.def
   \def\N{{\mathbb N}}
   
\begin{document}
\voffset=-0.5cm
\ytableausetup{boxsize=1.25em}

\title[]{Hopping in the Phase Model to a Non-Commutative Verlinde Formula for Affine Fusion}

\author{M.A. Walton}

\address{Department of Physics and Astronomy, University of Lethbridge, Lethbridge, Alberta, T1K 3M4, Canada\ }
\ead{walton@uleth.ca}
\begin{abstract}
Korff and Stroppel discovered a  realization of $su(n)$ affine fusion, the fusion of the $su(n)$ Wess-Zumino-Novikov-Witten (WZNW) conformal field theory, in the phase model, a limit of the $q$-boson hopping model. This integrable-model realization provides a new perspective on affine fusion, explored in a recent paper by the author.  The role of WZNW primary fields is played in it by non-commutative Schur polynomials, and fusion coefficients are thus given by a non-commutative version of the Verlinde formula.  We present the extension to all Verlinde dimensions, of  arbitrary genus and any number $N$ of points. The level-dependence of affine fusion is also discussed, using the concept of threshold level, and its generalization to threshold weight.
\end{abstract}

\date{\today}

\section{Introduction}

Affine fusion is  found in several  mathematical and physical contexts. It is a natural generalization of the tensor product of representations of simple Lie algebras; a simple truncation thereof controlled by a non-negative integer, the level.

One important physical context is conformal field theory, and more specifically, the Wess-Zumino-Novikov-Witten (WZNW) models (see \cite{DMS}, for example). WZNW models realize, at a fixed non-negative integer level $k$, a non-twisted affine Kac-Moody algebra $g^{(1)}$ based on a simple Lie algebra $g$, or $g_k$ for short. Their primary fields furnish representations of $g_k$ and their operator products are governed by the corresponding affine fusion algebra.

Recently, Korff and Stroppel \cite{KS} found a much simpler physical realization of affine fusion, for the $su(n)_k$ case, at all levels $k\in \N$ together (see also \cite{K, MW}).  The phase model \cite{BIK} is an integrable multi-particle model whose integrals of motion are non-commutative Schur polynomials \cite{KS}.  The integrals may be diagonalized by the algebraic Bethe ansatz, and their eigenvalues are affine fusion eigenvalues \cite{KS}.  The non-commutative Schur polynomials play the role in the phase model that the primary fields do in WZNW models \cite{MW}. The correlators of non-commutative Schur polynomials equal affine fusion coefficients \cite{KS} and more generally, affine Verlinde dimensions of arbitrary genus and numbers of marked points \cite{MW}. The Korff-Stroppel formula for these correlators can be thought of as a non-commutative Verlinde formula.

This proceedings contribution is based mostly on the paper \cite{MW}. Section 2 reviews the phase model realization of affine fusion found by Korff and Stroppel \cite{KS}. Section 3 records the new results of \cite{MW}: the extension of the non-commutative Verlinde formula to higher genus and $N$ points, and consideration of the level-dependence of fusion in the phase-model realization. In the latter, the old concept of threshold level \cite{CMW,KMSW} plays a prominent role. Generalization to consideration of threshold weights \cite{MWcjp} is included here as well.  Section 4 is a short conclusion.


\section{Phase-Model Realization of Affine ${\mathbf su(n)}$ Fusion}

In the $su(n)$ phase model, bosons  occupy $n$ sites on $S^1$, or, equivalently, the nodes of the Coxeter-Dynkin diagram of the affine Kac-Moody algebra $\widehat{su}(n)\cong A^{(1)}_{n-1}$. For each $j\in\{1, \ldots,n\}$, the $j$-th node is associated to an  affine fundamental weight $\Lambda^j$, and to boson creation and annihilation operators $\vpd\sub{j}$ and $\vp\sub{j}$, respectively. The affine dominant weight $\hat\nu\ =\ \sum\sub{\, j=1}^n\, \nu\sub{j}\, \Lambda^j$ can be used to label a basis of states, where $\nu\sub{j}\in\N_0$ is the number of bosons at node $j$, eigenvalue of the number operator $N_j$. The complete set of possible affine dominant weights is $\hat P_+ := \big\{  \sum_{a=1}^{n} \nu_a \Lambda^a  \,\vert\,  \lambda_a\in \N_0 \big\}$.

The total number of bosons $\sum\sub{\, j=1}^n\, \nu\sub{j}$ equals the level of $\hat\nu$, denoted $k$. So, alternatively, the level $k$ and an $su(n)$ dominant weight $\nu = \sum\sub{\, j=1}^{n-1} \nu\sub{j} \Lambda^j$ can be used to label states, since $\nu\sub{n} = k- \sum\sub{\, j=1}^{n-1}\nu\sub{j}$.  Notice that $\nu\sub{n}$ is the affine Dynkin label. The 2  notations are useful, so we write $\vert\hat\nu\rangle\, =\, \vert\nu\rangle_k$. Here
$\hat\nu \in  \hat P_+^k := \big\{  \sum_{a=1}^{n} \nu_a \Lambda^a  \,\vert\,  \lambda_a\in \N_0,\, \sum_{a=1}^{n} \nu_a = k \big\}\, \subset\, \hat P_+$,
and $\nu \in  P_+^k  := \Big\{  \sum_{a=1}^{n-1} \nu_a \Lambda^a\,  \vert\,  \lambda_a\in \N_0,\, \sum_{a=1}^{n-1} \nu_a \leq k \big\}$. The basis vectors are orthonormal, so \begin{equation}  {}_k\langle \lambda\vert \mu\rangle_{k'}\ =\ \delta_{\lambda,\mu}\, \delta_{k,k'}\ \ ,\, {\rm or}\ \ \ \langle\hat\lambda \vert\hat\mu \rangle\ =\ \delta_{\hat\lambda, \hat\mu}\ \ .\label{orth}\end{equation} It is important to realize, however, that the level, the number of bosons, is not fixed in the phase model.

The algebra of $\{\,\vp_j, \vpd_j, N_j \,|\, j\in\{1,\ldots,n\}\, \}$, subject to the following relations, is known as the {\it phase algebra}.
\begin{equation*} [\vp\sub{i},\vp\sub{j}]\ =\ 0\ ,\ \ \   [\vpd\sub{i}, \vpd\sub{j}]\ =\ 0\ ,\ \ \   [N_i, N_j]\ =\ 0\ , \end{equation*}
\begin{equation*} [N_i, \vpd\sub{j}]\ =\ \delta_{i,j}\,\vpd\sub{i}\ , \ \ \  [N_i, \vp\sub{j}]\ =\ -\delta_{i,j}\,\vp\sub{i}\ ,\quad \end{equation*}
\begin{equation*} N_i\,(1-\vpd\sub{i}\vp\sub{i})\ =\ 0\ =\ (1-\vpd\sub{i}\vp\sub{i})\, N_i\ ,\end{equation*}
\begin{equation} [\vp\sub{i},\vpd\sub{j}]\ =\ 0\ \ \  {\rm if}\ i\not=j\ ; \ \ {\rm but}\ \ \  \vp\sub{i}\vpd\sub{i}\ =\ 1\ .\label{phase}\end{equation}
The last relation follows from the standard actions of $\vp_i$ and $\vpd_i$ on the basis states:
\begin{equation*} \vp_i\,\vert\hat\nu\rangle\ =\ \left\{\begin{array}{cc} \vert\hat\nu-\Lambda^{i}\rangle\ \ ,
\ & \ \ \hat\nu-\Lambda^{i}\in \hat P_+\ ;\cr  0\ ,\ & \ {\rm otherwise}\ .\end{array}\right.  \end{equation*}
\begin{equation*} \vpd_i\,\vert\hat\nu\rangle\ =\  \vert\hat\nu+\Lambda^{i}\rangle\ \ .
  \end{equation*}
Similarly, the operator $\vpd\sub{i}\vp\sub{i}$ can be seen to project onto states of positive $i$-th boson number $\nu_i>0$, so that $1-\vpd\sub{i}\vp\sub{i}$ projects onto $\nu_i=0$ states.

The phase model is solved in the standard way. The Lax matrices are \begin{equation}L_j(u)\ =\ \left(\matrix{ 1 & u\vpd\sub{j}\cr    \vp\sub{j} & u }\right)\, ,\label{Lax}\end{equation} where $u$ is a spectral parameter. The monodromy matrix is then \begin{equation}  M(u)\ =\ L_n(u)\, L_{n-1}(u)\,\cdots\, L_1(u)\ =:\ \left(\begin{array}{cc} A(u) & B(u) \\ C(u) & D(u) \end{array}\right)\ ,  \label{mono}\end{equation} where the last equality is just standard notation.  Integrability results because the fundamental relation \begin{equation} R_{12}(u/v)\, M_1(u)\, M_2(v)\ =\ M_2(v)\, M_1(u)\, R_{12}(u/v)\ ,\label{RMM}\end{equation} is satisfied.  This guarantees that the
quantum Yang-Baxter equation \begin{equation*} R_{12}(u/v)\, R_{13}(u)\, R_{23}(v)\ =\ R_{23}(v)\, R_{13}(u)\, R_{12}(u/v)\ \end{equation*} follows, here with $R$-matrix
\begin{equation*} R(x)\ =\ \left(\begin{array}{cccc} \frac x {x-1} & 0 & 0 & 0 \\ 0 & 0 &  \frac x {x-1} & 0 \\ 0 & \frac 1 {x-1} & 1 & 0 \\ 0 & 0 & 0 & \frac x {x-1} \end{array}\right)\ .\end{equation*}

The relation (\ref{RMM}) defines the so-called  Yang-Baxter algebra, that includes the vanishing commutator
$[B(u), B(v)]\, =\, 0$.  The algebraic Bethe ansatz then yields   $B(x_1)\cdots B(x_k)\,\vert \hat 0 \rangle$ for the Bethe vectors. Notice that the latter form is symmetric in the commuting creation operators $B(x_j)$, and so completely symmetric in the variables $x_1, \ldots, x_k$. As a result, it can be expanded in terms of the symmetric Schur polynomials:  \begin{equation*}  B(x_1)\cdots B(x_k)\,\vert \hat  0 \rangle\ =\ \sum_{\lambda\in P_+^k}\, s\sub{\lambda^t} (x_1, \ldots, x_k)\,\vert \lambda \rangle\sub{k}\ \ .\end{equation*}  The Schur symmetric polynomial $s\sub{\lambda^t} (x_1, \ldots, x_k)$ is also an  $su(n)$ character.

The Bethe vectors diagonalize the integrals of motion, provided the Bethe ansatz equations are satisfied, and the latter imply that $x_1,\ldots,x_k$ must equal certain $(k+n)$-th roots of unity. More specifically, one can put the possible sets of values of the $x_j$ in 1-1 correspondence with dominant weights in $P_+^k$.  Adopting the notation $(x_1,\ldots,x_k) =: x\sub{\sigma}$ for $\sigma\in P_+^k$, one finds that $s\sub{\lambda}(x\sub{\mu})$ is an affine fusion eigenvalue for $\sigma\in P_+^k$. The celebrated Verlinde formula
 \begin{equation}  s\sub{\lambda}(x\sub{\sigma})\, s\sub{\mu}(x\sub{\sigma})\ =\ \sum_{\nu\in P_+^k}\, {}^{(k)}N_{\lambda,\mu}^\nu\, s\sub{\nu}(x\sub{\sigma})\ \label{Vs}\end{equation}
follows.  Here $\lambda, \mu, \nu, \sigma \in P_+^k$ and  ${}^{(k)}N_{\lambda,\mu}^\nu$ is an affine fusion multiplicity, and the connection with WZNW models is made.

Affine fusion multiplicities count the couplings of primary fields in a WZNW model. Affine fusion is a natural generalization of the tensor product of representations of simple Lie algebras. More precisely, it is a simple truncation thereof, controlled by the level $k$.  If $T_{\lambda,\mu}^\nu$ indicates the $su(n)$ tensor product multiplicity, then  \begin{equation*}  {}^{(k)}N_{\lambda,\mu}^\nu\ \leq\  T_{\lambda,\mu}^\nu\ ;\ \qq\  {}^{(\infty)}N_{\lambda,\mu}^\nu\ =\  T_{\lambda,\mu}^\nu\ ,  \end{equation*}  for all $\lambda, \mu, \nu \in P^k_+$.

The direct connection with affine fusion is revealed by examining the integrals of motion \cite{KS}.  Since fusion has fixed level $k$, we consider hopping operators $a\sub{j}\ =\ \vpd\sub{j}\,\vp\sub{j-1}$, $j\in\{1,2,\ldots,n\}$, that do not change the number of bosons. For $\hat\nu\in \hat P_+^k$, their action, depicted in Fig. 1, is
\begin{equation*} a_i\,\vert\hat\nu\rangle\ =\ \left\{\begin{array}{cc} \vert\hat\nu-\Lambda^{i-1}+\Lambda^{i}\rangle\ \ ,
\ & \ \ \hat\nu-\Lambda^{i-1}+\Lambda^{i}\in \hat P_+^k\ ;\cr  0\ ,\ & \ {\rm otherwise}\ .\end{array}\right.  \end{equation*}

\begin{figure}[ht]
\begin{center}
\begin{tikzpicture}[auto,thick,main node/.style={circle,draw,font=\sffamily\small\bfseries}]
 \coordinate (N1) at (90: 2cm);
 \coordinate (N2) at (135: 2cm);
 \coordinate (N3) at (180: 2cm);
 \coordinate (N4) at (225: 2cm);
 \coordinate (N5) at (270: 2cm);
 \coordinate (N6) at (315: 2cm);
 \coordinate (N7) at (0: 2cm);
 \coordinate (N8) at (45: 2cm);
 \node[main node] (1) at (N1) {$\nu\sub{1}$};
 \node[main node] (2) at (N2) {$\nu\sub{2}$};
 \node[main node] (3) at (N3) {$\nu\sub{3}$};
 \node[main node] (4) at (N4) {$\nu\sub{4}$};
 \node[main node] (5) at (N5) {$\nu\sub{5}$};
 \node[main node] (6) at (N6) {$\nu\sub{6}$};
 \node[main node] (7) at (N7) {$\nu\sub{7}$};
 \node[main node] (8) at (N8) {$\nu\sub{8}$};
 \path
    (1) edge  (2)
    (2) edge  (3)
    (3) edge  (4)
    (4) edge  (5)
    (5) edge  (6)
    (6) edge  (7)
    (7) edge  (8)
    (8) edge  (1);
\path
(6)   edge[->,densely dotted, shorten <=3pt,
           shorten >=3pt, bend left,{font=\sffamily\large\bfseries}] node[left] {$a\sub{7}$} (7);
\end{tikzpicture}
\end{center}
\caption{The action of the hopping operator $a\sub{7}$ is indicated in the $su(n)=su(8)$ phase model.}
\label{figure:suEight}
\end{figure}
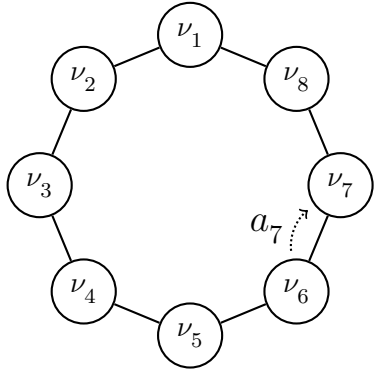

The algebra of  hopping operators $\cA=\langle a_1,a_2,\ldots,a_n\rangle$ is given by:
\begin{equation*}  [a_i,a_j]\ =\ 0\, ,\ \ \ \ {\rm if}\  i\not = j\pm 1\ {\rm mod}\ n\ ;\end{equation*}
\begin{equation} a_{i}a_j^{\, 2}\ =\ a_j a_{i} a_j\ \ ,\  \ \    a_{i}^{\, 2}a_j\ =\ a_{i} a_j a_{i}\ \ \ \  {\rm if}\ i=j+1\ {\rm mod}\ n\ . \label{cA}\end{equation}
Here indices are defined mod $n$. This hopping algebra $\cA$ is called the local affine plactic algebra by Korff and Stroppel \cite{KS}. The term plactic indicates a connection with Young tableaux (see \cite{F}, e.g.). $su(n)$ tensor products can be calculated using Young tableaux in the famous Littlewood-Richardson rule. The integrals of motion of the phase model led Korff and Stroppel to a modified Littlewood-Richardson rule for affine fusion \cite{KS}.

To see the relation to Young tableaux, note that the hopping operator $a\sub{i}$  is associated with the weight $\Lambda^{i}-\Lambda^{i-1}$. The set of these affine weights have horizontal parts equal to the weights of the basic $su(n)$ irreducible representation $L(\Lambda^1)$, that can be labelled by Young tableau $\begin{ytableau}{i}\end{ytableau}\,$.  For example, in the $su(3)$ case, the tableaux for $L(\Lambda^1)$ are
\begin{equation*}  \ \begin{ytableau}2 \end{ytableau}\ \quad\ \ \ \begin{ytableau}1 \end{ytableau}\qquad \ . \end{equation*}
\begin{equation*} \qquad \begin{ytableau}3\end{ytableau}\end{equation*}
They have been arranged here in the pattern of the weight diagram for $L(\Lambda^1)$.

Furthermore, the action of hopping operators on basis states $\vert\hat\lambda\rangle\, ,\ \hat\lambda\in \hat P_+^k$, is
\begin{equation*} a_i\,\vert\hat\lambda\rangle\ =\ \left\{\begin{array}{cc} \vert\tilde\lambda-\Lambda^{i-1}+\Lambda^{i}\rangle\ \ ,
\ & \ \ \hat\lambda-\Lambda^{i-1}+\Lambda^{i}\in \hat P_+^k\ ;\cr  0\ ,\ & \ {\rm otherwise}\ .\end{array}\right.  \end{equation*}
It is precisely reproduced by the product (denoted by $*$ here) of a Young tableaux and a Young diagram that is used in the Littlewood-Richardson rule. In the $su(3)$ example:
\begin{equation*} \begin{ytableau}2 \end{ytableau}\, {*\atop{\ }}\, \begin{ytableau}{\ }& {\ }\end{ytableau}\, {=\atop{\ }}\, \begin{ytableau}{\ }&{\ }\\ 2\end{ytableau}\ {,\atop{\ }}\quad\qquad \begin{ytableau}1 \end{ytableau}\, {*\atop{\ }}\, \begin{ytableau}{\ }& {\ }\end{ytableau}\, {=\atop{\ }}\, \begin{ytableau}{\ }&{\ }& 1\end{ytableau}\ {,\atop{\ }}\end{equation*}
\begin{equation*} \begin{ytableau}3 \end{ytableau}\, {*\atop{\ }}\, \begin{ytableau}{\ }& {\ }\end{ytableau}\ {=\atop{\ }}\ \begin{ytableau}{\ }&{\ }\\ \none\\ 3\end{ytableau}\ {=\atop{\ }}\ {0\ .\atop{\ }}\ \end{equation*}
The last result vanishes because the resulting tableau has non-dominant shape.

Of course, more complicated Young tableaux must also be considered. Take the adjoint representation of $su(3)$, of highest weight $\Lambda^1 +\Lambda^2$. Its Young tableaux are
\begin{equation*}  \qquad\qquad\qquad\begin{ytableau}1 & 2\\ 2\end{ytableau}\ \qquad\ \ \qquad \begin{ytableau}1 & 1\\ 2\end{ytableau} \end{equation*}
\vskip.5cm\begin{equation*} \begin{ytableau}2 & 2\\ 3\end{ytableau}\qquad\qquad\quad   \begin{ytableau}1 & 3\\ 2\end{ytableau}\ \begin{ytableau}1 & 2\\ 3\end{ytableau}\qquad\qquad \quad   \begin{ytableau}1 & 1\\ 3\end{ytableau}\end{equation*}
\vskip.5cm\begin{equation*} \qquad\qquad\qquad\begin{ytableau}2 & 3\\ 3\end{ytableau}\ \qquad\ \ \qquad \begin{ytableau}1 & 3\\ 3\end{ytableau}  \end{equation*}
As a more non-trivial example of the Littlewood-Richardson rule, the Young tableaux above may be added column-by-column, from right to left, to the Young diagram $\,\begin{ytableau} {\ }\end{ytableau}$ to compute the $su(3)$ tensor product decomposition
\begin{equation*} L(\Lambda^1+\Lambda^2)\otimes L(\Lambda^1)\ \hookrightarrow\ L(2\Lambda^1+\Lambda^2)\oplus L(2\Lambda^2) \oplus L(\Lambda^1)\ .\end{equation*}
Restricting attention to the 2 weight-0 tableaux, we find no contribution and
$L(\Lambda^1+\Lambda^2)\otimes L(\Lambda^1)\,\supset\, L(\Lambda^1)$, from \begin{equation*} \begin{ytableau}1 & 3\\ 2\end{ytableau}\ {*\atop{\ }}\ \begin{ytableau}{\ }\end{ytableau}\ \Rightarrow\ \  \begin{ytableau}{\ }\\ \none \\ 3\end{ytableau}\ \ \ \qquad{\rm and}\end{equation*}
\begin{equation*} \begin{ytableau}1 & 2\\ 3  \end{ytableau}\ {*\atop{\ }}\ \begin{ytableau}{\ }\end{ytableau}\ \Rightarrow\ \  \begin{ytableau}{\ }\\ 2\end{ytableau}\ \ ,\ \ \ \begin{ytableau}{\ }& 1\\ 2\\ 3\end{ytableau}\   \ \ ,\end{equation*}
respectively.

For successive tensor products, the algorithm can be repeated.  Alternatively, a product $\bullet$ of 2 Young tableaux can be defined.  One finds \cite{F} that for consistency, certain ``bumping'' relations must be obeyed:
\begin{equation*} \begin{ytableau}
j & k
\end{ytableau}\ {\bullet\atop{ }}\  \begin{ytableau} i\end{ytableau}\ {=\atop{}}\ \begin{ytableau}
i & k \\
 j
\end{ytableau}\ \ ,\ \ \ i<j\le k\ ;\end{equation*}
\begin{equation} \begin{ytableau}
i & k
\end{ytableau}\ {\bullet\atop{}}\  \begin{ytableau} j\end{ytableau}\ {=\atop{}}\ \begin{ytableau}
i & j \\
 k
\end{ytableau}\ \ ,\ \ \ i\le j<k\ .\label{bump}\end{equation}

All such computations can be done in plactic algebras, instead of with tableaux. First, replace Young tableaux with  words. The (column) word of a Young tableau is obtained by listing its entries in the order from bottom to top in the left-most column, then from bottom to top in the next-to-left-most column, continuing until the top entry of the right-most column is listed. The (column) words for the adjoint representation of $su(3)$, of highest weight $\Lambda^1 +\Lambda^2$, are
\begin{equation*}  \quad\qquad\qquad 2\,1\,2\ \qquad\ \ \qquad 2\,1\,1 \end{equation*}
\begin{equation*}  3\,2\,2\qquad\qquad\quad   2\,1\,3\ \ \ \ 3\,1\,2\qquad\qquad \quad   3\,1\,1\qquad .\end{equation*}
\begin{equation*} \quad\qquad\qquad 3\,2\,3\ \qquad\ \ \qquad 3\,1\,3  \ ,\end{equation*} for example.
Replacing each occurrence of digit $j$ in the word by the corresponding generator of the plactic algebra  ${\bar a}\sub{j}$ yields
\begin{equation*}  \ \ \quad\qquad\qquad {\bar a}_2{\bar a}_1{\bar a}_2\ \qquad\ \ \qquad {\bar a}_2{\bar a}_1{\bar a}_1 \end{equation*}
\begin{equation} {\bar a}_3{\bar a}_2{\bar a}_2\qquad\qquad\quad   {\bar a}_2{\bar a}_1{\bar a}_3\ \ \ \ {\bar a}_3{\bar a}_1{\bar a}_2\qquad\qquad \quad   {\bar a}_3{\bar a}_1{\bar a}_1\quad .\label{badj}\end{equation}
\begin{equation*} \ \ \quad\qquad\qquad {\bar a}_3{\bar a}_2{\bar a}_3\ \qquad\ \ \qquad {\bar a}_3{\bar a}_1{\bar a}_3 \ .  \end{equation*}

Most important are relations (\ref{bump}) re-written as
\begin{equation*} {\bar a}_{i+1}^2{\bar a}_i = {\bar a}_{i+1}{\bar a}_i{\bar a}_{i+1}\qquad\quad (j=k=i+1)\ ;\end{equation*}
\begin{equation} {\bar a}_i{\bar a}_{i+1}{\bar a}_i = {\bar a}_{i+1}{\bar a}_i^2\qquad\quad (j=k=i+1)\ .\label{bcA}\end{equation}
Comparing (\ref{bcA}) with (\ref{cA}), we see that bumping is compatible with hopping. That is, the relations defining the hopping algebra (\ref{cA}) are very similar to those resulting from the bumping process in (\ref{bump}). The algebra of hopping operators can therefore implement calculations similar to those involved in the Littlewood-Richardson rule for $su(n)$ tensor products.  Korff and Stroppel showed that the hopping operators of the phase model realize $su(n)$ affine fusion, a truncation of the $su(n)$ tensor product.

The integrals of motion can be described explicitly. The fundamental relation (\ref{RMM})
guarantees that the transfer matrix $T(u)\ :=\ {\rm tr}\, M(u)$ obeys
\begin{equation}\big[\,T(u), T(v)\,\big]\ =\ 0\ ,\label{TT}\end{equation}
and so is the generating function \begin{equation*} T(u)\ =\ \sum_{r=0}^n\, u^r\, e_r(\cA)\  \end{equation*} of integrals of motion \begin{equation*} [\, e_r(\cA),\, e_{r'}(\cA)\, ]\ =\ 0\ \ .\end{equation*}

The form of the integrals can be found from (\ref{Lax},\ref{mono}). $e_r(\cA)$ indicates the $r$-th cyclic elementary symmetric polynomial, the sum of all cyclically ordered products of $r$ distinct hopping operators $a_i$: \begin{equation*} e_r(\cA)\ =\ \sum_{\vert I\vert=r}\,\prod_{i\in I}^{\circlearrowright}\, a_i\ \ .\end{equation*} For example, with $n=4$, \begin{equation*} e_2(\cA)\ =\ a_2a_1+a_3a_1 +a_1a_4+a_3a_2 +a_4a_2+a_4a_3\ .\end{equation*}

Because of the integrability property (\ref{TT}), one can use the Jacobi-Trudy formula to define the  non-commutative Schur polynomial \begin{equation}  s_\lambda(\cA)\ =\ \det\left( \,e_{\lambda_i^t-i+j}(\cA)\,  \right)\ .    \label{JT}\end{equation}
Here $\lambda^t$ indicates the partition specified by the transpose of the Young diagram for $\lambda$. $\lambda_i^t$ is the $i$-th integer in that partition.

For example, the non-commutative Schur polynomial $s\sub{\Lambda^1+\Lambda^2}(\cA)$ for the adjoint  representation of $su(3)$ equals the sum of:
\begin{equation*}  \ \ \quad\qquad\qquad a_2a_1a_2\ \qquad\ \qquad a_1a_2a_1 \end{equation*}
\begin{equation} a_2a_3a_2\qquad\qquad\quad\   {{(a_3a_2a_1+a_1a_3a_2}\atop{+a_2a_1a_3-1)}} \qquad\qquad \quad\   a_1a_3a_1\quad .\label{adj}\end{equation}
\begin{equation*} \ \ \quad\qquad\qquad a_3a_2a_3\ \qquad\  \qquad a_3a_1a_3  \end{equation*}
Comparing this last result with (\ref{badj}), we already see a couple of important differences. First, (\ref{adj}) contains negative terms, whereas (\ref{badj}) contains none. This is general: the vectors of any representation $L(\lambda)$ of $su(n)$ can be put in 1-1 correspondence with Young tableaux.  But the corresponding non-commutative Schur polynomial is a sum that includes negative terms, in general. Second, there is a cyclic symmetry present in (\ref{adj}) that is broken in (\ref{badj}).

The main result of \cite{KS} is
\begin{equation} s\sub{\lambda}(\cA)\, \vert\, \mu\,\rangle\sub{k}\ =\ \sum_{\nu\in P_+^k}\, {}^{(k)}N_{\lambda,\mu}^\nu\, \vert \, \nu\,\rangle\sub{k}\ \ .\label{ncV}\end{equation}
Inspired by the term non-commutative Schur polynomial, one can call the Korff-Stroppel equation (\ref{ncV}) a non-commutative Verlinde formula. Using the orthonormality (\ref{orth}), it can be written as
\begin{equation*}  {}\sub{k}\langle\,\nu\,\vert\, s\sub{\lambda}(\cA)\, \vert\,\mu\,\rangle\sub{k}\ =\  {}^{(k)}N_{\lambda,\mu}^\nu\ \ .\end{equation*}

To compare with  WZNW conformal field theory, consider a special case of (\ref{ncV}), $ s\sub{\lambda}(\cA)\vert  0 \rangle\sub{k} =  \vert \lambda \rangle\sub{k}\ .$ This is highly reminiscent of the field-state correspondence   $\phi\sub{\lambda}(0) \vert 0\rangle  =  \vert \lambda \rangle ,$ involving the WZNW primary field $\phi_\lambda(z)$. Furthermore, it can be used to re-write (\ref{ncV}) as
\vskip0.01cm
\begin{minipage}[b]{0.5\textwidth}
\begin{tabbing}
 \\
${}\ \ \ {}^{(k)}N_{\lambda,\mu,\nu}\ =\ {}\sub{k}\langle 0 \vert\, s\sub{\lambda}(\cA)\, s\sub{\mu}(\cA)\, s\sub{\nu}(\cA)\, \vert 0 \rangle\sub{k}\ \ =\ $\\ \\
\end{tabbing}
\end{minipage}
\hspace{.25cm}
\begin{minipage}[b]{0.25\textwidth}
\centering
\tikzset{
    photon/.style={decorate, decoration={snake}, draw=red},
    electron/.style={draw=black, postaction={decorate},
        decoration={markings,mark=at position .55 with {\arrow[draw=black]{>}}}},
    positron/.style={draw=black, postaction={decorate},
        decoration={markings,mark=at position .55 with {\arrow[draw=black]{<}}}},
    gluon/.style={decorate, draw=magenta,
        decoration={coil,amplitude=4pt, segment length=5pt}}
}
\begin{tikzpicture}[thick,scale=0.5,
        level/.style={level distance=1.5cm},
        level 2/.style={sibling distance=2.6cm},
        level 3/.style={sibling distance=2cm}]
    \coordinate
        child[grow=left]{
            child {
                node {$\lambda$}
                edge from parent [positron]
            }
            child {
                node {$\mu$}
                edge from parent [positron]
            }
            edge from parent [electron] node [above right=4pt] {$\nu$}
        }
        ;
\end{tikzpicture}
\end{minipage}
\vskip0.01cm\noindent where the trivalent, directed, labeled graph that normally indicates the fusion coefficient has been drawn. Compare the last expression with the notation for the 3-point function in WZNW model: $\langle 0\vert\, \phi_\lambda(z_1)\, \phi_\mu(z_2)\, \phi_\nu(z_3)\, \vert 0\rangle\ .$ It becomes clear that the non-commutative Schur polynomial $s\sub{\lambda}(\cA)$ plays the role in the phase model of the primary field $\phi\sub{\lambda}$ in the WZNW conformal field theory.
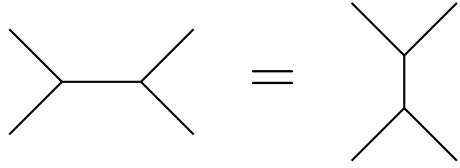
\begin{figure}[ht]
\begin{center}
\begin{tikzpicture}[scale=0.35,
        thick,
        level/.style={level distance=1.5cm},
        level 2/.style={sibling distance=2.6cm},
        level 3/.style={sibling distance=2cm}
    ]
 \coordinate (N1) at (-10cm,2cm);
 \coordinate (N2) at (-10cm,-2cm);
 \coordinate (N3) at (-8cm,0cm);
 \coordinate (N4) at (-5cm,0cm);
 \coordinate (N5) at (-3cm,2cm);
 \coordinate (N6) at (-3cm,-2cm);
 \coordinate (N7) at (3cm,3cm);
 \coordinate (N8) at (7cm,3cm);
 \coordinate (N9) at (5cm,1cm);
 \coordinate (N10) at (5cm,-1cm);
 \coordinate (N11) at (3cm,-3cm);
 \coordinate (N12) at (7cm,-3cm);
\draw[thick] (N3) -- (N4);
\draw[thick] (N1) -- (N3) -- (N2);
\draw[thick] (N5) -- (N4) -- (N6);
\draw[thick] (N9) -- (N10);
\draw[thick] (N7) -- (N9) -- (N8);
\draw[thick] (N11) -- (N10) -- (N12);
\draw[thick] node at (0,0) {{\huge$=$}};
\end{tikzpicture}
\end{center}
\caption{The duality of affine fusion is a consequence of WZNW duality. In the phase model, affine fusion duality is a result of integrability.}
\label{figure:duality}
\end{figure}

Remarkably, it is the integrability of the phase model that gives rise to duality in affine fusion. The non-commutative Schur polynomials are integrals of motion, and so commute: $[ s\sub{\lambda}, s\sub{\mu} ] = 0$. Here we have used $s_\lambda\ :=\ s_\lambda(\cA)$, for short. Consequently,
\begin{equation*} {}\sub{k}\langle 0 \vert\, s\sub{\lambda}\, s\sub{\mu}\, s\sub{\nu}\, s\sub{\phi}\, \vert 0 \rangle\sub{k}\ =\ \QQ\qq\end{equation*}
\begin{equation*} \ \ \sum_{\sigma\in P_+^k}\, {}\sub{k}\langle 0 \vert\, s\sub{\lambda}\, s\sub{\mu}\,\vert \sigma \rangle\sub{k}\ \  {}\sub{k}\langle \sigma\vert\, s\sub{\nu}\, s\sub{\phi}\, \vert 0 \rangle\sub{k} \ =\
\ \ \sum_{\sigma\in P_+^k}\, {}\sub{k}\langle 0 \vert\, s\sub{\lambda}\, s\sub{\phi}\,\vert \sigma \rangle\sub{k}\ \  {}\sub{k}\langle \sigma\vert\, s\sub{\mu}\, s\sub{\nu}\, \vert 0 \rangle\sub{k}\ .
\end{equation*}
The graphical representation of this result is shown in Fig. \ref{figure:duality}.

\section{Higher-Genus and Level-Dependence}

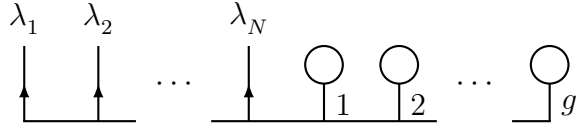
\begin{figure}[ht]
\begin{center}
\begin{tikzpicture}[thick, x=1cm, y=1cm, ray/.style={decoration={markings,mark=at position .5 with {
      \arrow[>=latex]{>}}},postaction=decorate}]
\draw[ray] (-3.985,0)--(-3.985,1.0) node[above] {$\lambda\sub{1}$};
\draw(-4, 0) -- (-2.5,0);
\draw (-2,0.25) node[above] {$\cdots$};
\draw (-1.5,0) -- (1.5,0);
\draw[ray](-3,0)--(-3,1.0) node[above] {$\lambda\sub{2}$};
\draw[ray](-1,0)--(-1,1.0) node[above] {$\lambda\sub{N}$};
\draw(0,0)--(0,0.5) node[below right] {$1$};
\draw(0,0.75) circle (0.25) ;
\draw(1,0)--(1,0.5) node[below right] {$2$};
\draw(1,0.75) circle (0.25);
\draw (2.0,0.25) node[above] {$\cdots$};
\draw(2.5,0) --(3.015,0);
\draw(3,0)--(3,0.5) node[below right] {$g$};
\draw(3,0.75) circle (0.25);
%
\end{tikzpicture}
\end{center}
\caption{The Verlinde dimension of arbitrary genus $g$ and $N$ points.}
\label{figure:Vdim}
\end{figure}

Let us now turn to the new results reported in \cite{MW}. First, consider higher-genus affine fusion. The Verlinde dimension of arbitrary genus and number of points is indicated by its trivalent graph in Fig. \ref{figure:Vdim}.  It is not difficult to show that the Korff-Stroppel non-commutative Verlinde formula extends to
\begin{equation*} {}^{(k,g)}N_{\lambda_1,\dots,\lambda_N}\ =\   \langle \lambda_1^*\vert\, \bigg(\,\sum_{\alpha\in P_+^k}\, s_{\alpha^*} s_\alpha \,\bigg)^g\, s_{\lambda_2}\cdots s_{\lambda_{N-1}}\,\vert\lambda_N\rangle\ \end{equation*}
\begin{equation}\qquad =\ {}\sub{k}\langle 0\vert\, \bigg(\, \sum_{\alpha\in P_+^k}\, s_{\alpha^*} s_\alpha \,\bigg)^g\, s_{\lambda_1}\, s_{\lambda_2}\cdots s_{\lambda_{N}}\,\vert 0\rangle\sub{k}\ \label{ncVdim}\end{equation} for this case.  One can then identify the handle operator as $\sum_{\alpha\in P_+^k}\, s_{\alpha^*} s_\alpha$.

Now consider the simple dependence of affine fusion on the level $k$,  well-described by the concept of threshold level \cite{CMW,KMSW}. All possible fusion decompositions can be given simply by treating the level as a variable, and writing multi-sets of threshold levels as subscripts. Consider an example: the decomposition of an $su(3)$ tensor product may be written \bea L(\Lambda^1+\Lambda^2)^{\,\otimes 2} \, \hookrightarrow &\, L(0)_2\,\oplus\,  2\,L(\Lambda^1+\Lambda^2)_{2, 3}\,\oplus\,  L(3\Lambda^1)_3\nn &\ \oplus L(3\Lambda^2)_3\,\oplus\,  L(2\Lambda^1+2\Lambda^2)_4\ .\quad\label{fpadjadj}\eea
The subscripts indicate the threshold levels of the representations in the decomposition, so that  \ben {}^{(2)}N\sub{\Lambda^1+\Lambda^2, \,\Lambda^1+\Lambda^2}^{\ \Lambda^1+\Lambda^2}\ =\ 1\ , \ \ {}^{(k\ge 3)}N\sub{\Lambda^1+\Lambda^2, \,\Lambda^1+\Lambda^2}^{\ \Lambda^1+\Lambda^2}\ =\ 2\ ,\een for example.

This threshold-level behaviour in affine fusion is reflected in the Korff-Stroppel result (\ref{ncV}) by the striking property of non-commutative Schur polynomials: $s\sub{\lambda}(\cA)$ does not depend on the level. The phase model treats all levels $k\in \N_0$ on an equal footing, and as a consequence, the threshold behaviour of level is clear. The level-dependence can be incorporated into  (\ref{ncV}) simply by using $\vert \mu\rangle\sub{k}$ with variable level.

In the $su(3)$  example just mentioned, the weight-0 part of the non-commutative Schur polynomial $s\sub{\Lambda^1+\Lambda^2}$ is
$a_3a_2a_1+a_1a_3a_2+a_2a_1a_3-1.$ But
\bea (a_3a_2a_1 & + a_1a_3a_2 +  a_2a_1a_3-1)\, \vert \Lambda^1+\Lambda^2+(k-2)\Lambda^3\rangle\nn &=\, \big\{ \,\theta(k-3)+\theta(k-2)\, \big\}\, \vert \Lambda^1+\Lambda^2+(k-2)\Lambda^3\rangle\,  ,\eea
confirming the presence of $L(\Lambda^1+\Lambda^2)_{2,3}$ in (\ref{fpadjadj}).
Here we have used $\vert \Lambda^1+\Lambda^2+(k-2)\Lambda^3\rangle = \theta(k-2) \vert \Lambda^1+\Lambda^2+(k-2)\Lambda^3\rangle$ and \ben \theta(x)\, :=\, \left\{ \begin{array}{cc} 1\ ,\ &\ \ {\rm if}\ x\ge 0\ ;\\ 0\ ,\ &\ {\rm if}\ \ x<0\ .    \end{array}\right.\ \een

As already emphasized, an advantage of the phase-model realization of affine fusion is that, unlike in the WZNW model, the level is not fixed--it is just the total particle number. Changes in level can be described in a simple, algebraic way by the operators $\vpd_i,\, \vp_i$ of the phase algebra (\ref{cA}).

Let us consider threshold multiplicities \cite{MW} in this spirit. The threshold multiplicity ${}^{(t)}n_{\lambda,\mu}^\nu$ is the contribution to a fusion multiplicity ${}^{(k)}N_{\lambda,\mu}^\nu$ at a fixed threshold level $t$, so that \ben {}^{(k)}N_{\lambda,\mu}^\nu\ =\  \sum_{t}^{k}\, {}^{(t)}n_{\lambda,\mu}^\nu\ \ .  \label{thmult}\een We also find \ben {}^{(k)}n_{\lambda,\mu}^\nu\ =\ {}^{(k)}N_{\lambda,\mu}^\nu\ -\  {}^{(k-1)}N_{\lambda,\mu}^\nu\ ,\label{nNN}\een where we have put ${}^{(k-1)}N_{\lambda,\mu}^\nu=0$ if any of $\lambda, \mu, \nu$ are not in $P_+^{k-1}$.

Notice that $\vpd_n\, \vert \mu\rangle_{k-1}\, =\, \vert \mu\rangle_{k}$. So we calculate \ben [s_\lambda(\cA), \vpd_n]\, \vert\mu\rangle_{k-1}\ =\ \sum_{\nu\in P_+^k} {}^{(k)}N_{\lambda, \mu}^\nu\, \vert \nu\rangle_k\, -\, \sum_{\nu\in P_+^{k-1}} {}^{(k-1)}N_{\lambda,\mu}^\nu\vp_n\, \vert \nu\rangle_{k-1}\ .\label{svp}\een So the phase-model version of (\ref{nNN}) is \ben {}_k\langle \nu\vert\,  [s_\lambda(\cA), \vpd_n]\, \vert \mu\rangle_{k-1}\ =\ {}^{(k)}n_{\lambda,\mu}^\nu\ \ . \label{slvpnn}\een
Once a particular non-commutative Schur polynomial $s_\lambda(\cA)$ is calculated, the interesting operator $[s_\lambda(\cA), \vpd_n]$ is easy to write down, since \ben  [a_i, \vpd_n]\ =\ \delta_{i,1}\, \vpd_1\, (1-\vpd_n\vp_n)\ .  \label{aivpn}\een
Recall that $1-\vpd_n\vp_n$ projects onto states with 0 particles at the $n$-th node.

To conclude this section, let us consider the threshold weight \cite{MWcjp}, a concept that generalizes threshold level, in the phase model. Instead of treating the level and the $n$-th Dynkin label as special, as in (\ref{ncV}), now let the weight $\hat\mu$ be variable in
\begin{equation} s\sub{\lambda}(\cA)\, \vert\, \hat\mu\,\rangle\ =\ \sum_{\nu\in \hat P_+^k}\, {}^{(k)}N_{\lambda,\mu}^\nu\, \vert \, \hat\nu\,\rangle\ , \ \ \ \ k=\sum_{c=1}^n\mu_c = \sum_{c=1}^n\nu_c\  .\label{ncVhat}\end{equation}
The existence of a threshold level $t$ implies a threshold value for the Dynkin label $\mu_n$, and threshold weight $\hat\theta$ will take account of all the $n$ Dynkin labels of $\hat\mu$.
A threshold weight $\hat\theta= \sum_{c=1}^n \theta_c \Lambda^c$ may be defined for each term $a\sub{j_L}\cdots a\sub{j_2}a\sub{j_1}$ in $s_\lambda(\cA)$. Put $\widehat{\rm wt}(a_j):= \Lambda^j-\Lambda^{j-1}$, $\widehat{\rm wt}(a_ia_j):= \widehat{\rm wt}(a_i) + \widehat{\rm wt}(a_j),$ etc. Then the $c$-th Dynkin label of the threshold weight is \ben  \theta_c(a\sub{j_L}\cdots a\sub{j_2}a\sub{j_1})\ =\ -{\rm min}\{ 0, {\rm wt}_c(a\sub{j_1}), \ldots, {\rm wt}_c(a\sub{j_L}\cdots a\sub{j_2}a\sub{j_1}) \,\}\ .\label{thwt}\een That is, $-\theta_c(a\sub{j_L}\cdots a\sub{j_2}a\sub{j_1})$ is the minimum value of the $c$-th Dynkin label of the weights of the sequence $a\sub{j_1}, a\sub{j_2}a\sub{j_1} ,\ldots, a\sub{j_L}\cdots a\sub{j_2}a\sub{j_1}$ and $0$.

For an $su(3)$ example, take  $\lambda=\Lambda^1+\Lambda^2$,\ and consider $a_3a_2a_1$ in  $s\sub{\lambda}(\cA)$.
The weight sequence associated with $a_3a_2a_1$ is $\{  {\widehat{\rm wt}}(a_1), {\widehat{\rm wt}}(a_2a_1),  \} = \{ \Lambda^1-\Lambda^3,  \Lambda^2-\Lambda^3, 0 \}\ .$
Therefore $\hat \theta(a_3a_2a_1) = \Lambda^3$.
This threshold weight tells us that $a_3a_2a_1$ contributes to $s_{\Lambda^1+\Lambda^2}(\cA)\,\vert \hat\mu\rangle$ iff   $\hat\mu\,  -\,  \hat\theta(a_3a_2a_1)\,=\, \hat\mu\, -\, \Lambda^3\, \in\, \hat P_+\ .$

\section{Conclusion}

The Korff-Stroppel integrable realization of $su(n)$ affine fusion \cite{KS} leads to the non-commutative Verlinde formula (\ref{ncV})  for affine fusion coefficients, that can be extended to a non-commutative formula (\ref{ncVdim}) for arbitrary Verlinde dimensions \cite{MW}. The realization offers a new perspective on affine fusion, that should deepen our understanding. It already makes clear certain features, like the existence of threshold levels (and weights) \cite{MW}.

Let us conclude with a brief discussion of possible future work.

Perhaps the most important further development would be to construct the phase-model realizations of affine fusion for all complex, simple Lie algebras. The next-to-simplest case is likely the Lie algebra $sp(2n)\cong C_n$, since tableaux work almost as well in this case as for $su(n)\cong A_{n-1}$.  Some progress is reported in \cite{AS}, where a quantum group approach is used.

From the physical point of view, it would be useful to derive the phase-model realization from affine fusion in another physical context. For example, it is well known that the $G/G$ gauged WZNW model is a topological field theory, with correlation functions equalling the affine Verlinde dimensions \cite{SYW}. Okuda and Yoshida \cite{OY} have already found the Bethe ansatz equations of the phase model in the path-integral formulation of the $U(n)/U(n)$ model, and other indications of the connection. A more direct relation would be helpful, as well as a manifestation of the phase model in the $su(n)$ Chern-Simons theory, which also has Verlinde dimensions as some of its expectation values \cite{W}.

More technically, formulas for $s\sub{\lambda}(\cA)$,  besides the Jacobi-Trudy formula (\ref{JT}), would likely prove useful. The properties of non-commutative Kostka polynomials should be explored.  Of course, similar formulas for non-commutative characters for other Lie algebras are desirable, too.

Finally, let us mention that non-negative integer matrix representations (NIM-reps) of fusion rings are important for boundary conformal field theories (see \cite{G}), e.g. the boundary WZNW model.  Can they be realized in a phase model context?

\vskip0.5cm
\noindent{\bf Acknowledgments}\hfb
The author thanks Ali Nassar and Andrew Urichuk for reading the manuscript. This research was supported in part by a Discovery Grant from the Natural Sciences and Engineering Research Council (NSERC) of Canada.


\vskip0.5cm
\section*{References}
\begin{thebibliography}{99}

\bibitem{AS} H.H. Anderson, C. Stroppel, Fusion rings for quantum groups, preprint(2012) [arXiv:1212.5736]

\bibitem{BIK} N.M. Bogoliubov, A.G. Izergin, N.A. Kitanine, Nucl. Phys. B 516 (1998) 501 [arXiv:solv-int/9710002]

\bibitem{CMW} C.J. Cummins, P. Mathieu, M.A. Walton, Phys. Lett. B 254 (1991) 386

\bibitem{DMS} P. Di Francesco, P. Mathieu, D. S\'en\'echal, Conformal Field Theory, (Springer-Verlag,
1997)

\bibitem{F} W. Fulton, Young Tableaux, London Math. Soc. Stud. Texts, vol. 35 (Cambridge U. Press, 1997)

\bibitem{G} T. Gannon, Nucl. Phys. B 627 (2002) 506 [arXiv:hep-th/0106105]

\bibitem{KMSW} A.N. Kirillov, P. Mathieu, D. S\'en\'echal, M.A. Walton, Nucl. Phys. B 391 (1993) 651 [arXiv:hep-th/9203004]

\bibitem{K} C. Korff, RIMS Kokyuroku Bessatsu B28 (2011) 121 [arXiv:1106.5342]

\bibitem{KS} C. Korff, C. Stroppel, Adv. Math. 225 (2010) 200 [arXiv:0909.2347]

\bibitem{OY}  S. Okuda, Y. Yoshida, JHEP 1211 (2012) 146 [arXiv:1209.3800]

\bibitem{SYW} M. Spiegelglas, S. Yankielowicz, Nucl. Phys. B 393 (1993) 301 [arXiv:hep-th/9201036] ; E. Witten, Commun. Math. Phys. 144 (1992) 189

\bibitem{V} E. Verlinde, Nucl. Phys. B 300 (1988) 360

\bibitem{W} E. Witten, Commun. Math. Phys. 121 (1989) 351

\bibitem{MW} M.A. Walton, SIGMA 8 (2012) 086 [arXiv:1208.0809]

\bibitem{MWcjp} M.A. Walton, Can. J. Phys. 72 (1994) 527









\end{thebibliography}


\end{document}